\documentclass[letterpaper, 12 pt, conference]{ieeeconf}  % Comment this line out
\IEEEoverridecommandlockouts                              % This command is only
\usepackage{graphicx} % for pdf, bitmapped graphics files
\usepackage{bm}
\usepackage{amsmath}
\usepackage[dvipsnames]{xcolor}
\usepackage{soul}
\usepackage{setspace} % for double spacing
\usepackage{soul}

\usepackage{amssymb}  % assumes amsmath package installed

\title{\LARGE \bf Dynamic Load Balancing for EV Charging Stations Using Reinforcement Learning and Demand Prediction
}

\author{Hesam Mosalli, Saba Sanami, Yu Yang,  Hen-Geul Yeh, and Amir G. Aghdam % <-this % stops a space
\thanks{Hesam Mosalli, Saba Sanami, and Amir G. Aghdam are with the
Department of Electrical and Computer Engineering, Concordia Univerity,
Montreal, QC, Canada. Emails: hesam.mosalli@mail.concordia.ca, saba.sanami@mail.concordia.ca,
amir.aghdam@concordia.ca.
Yu Yang is with the Department of Chemical Engineering and Hen-Geul Yeh is with the Department of  Electrical Engineering, California State University Long Beach. Emails: yu.yang@csulb.edu, henry.yeh@csulb.edu.}% <-this % stops a space
}

\begin{document}

\maketitle
\thispagestyle{empty}
\pagestyle{empty}

%%%%%%%%%%%%%%%%%%%%%%%%%%%%%%%%%%%%%%%%%%%%%%%%%%%%%%%%%%%%%%%%%%%%%%%%%%%%%%%%
\begin{abstract}
This paper presents a method for load balancing and dynamic pricing in electric vehicle (EV) charging networks, utilizing reinforcement learning (RL) to enhance network performance. The proposed framework integrates a pre-trained graph neural network to predict demand elasticity and inform pricing decisions. The spatio-temporal EV charging demand prediction (EVCDP) dataset from Shenzhen is utilized to capture the geographic and temporal characteristics of the charging stations. The RL model dynamically adjusts prices at individual stations based on occupancy, maximum station capacity, and demand forecasts, ensuring an equitable network load distribution while preventing station overloads. By leveraging spatially-aware demand predictions and a carefully designed reward function, the framework achieves efficient load balancing and adaptive pricing strategies that respond to localized demand and global network dynamics, ensuring improved network stability and user satisfaction. The efficacy of the approach is validated through simulations on the dataset, showing significant improvements in load balancing and reduced overload as the RL agent iteratively interacts with the environment and learns to dynamically adjust pricing strategies based on real-time demand patterns and station constraints. The findings highlight the potential of adaptive pricing and load-balancing strategies to address the complexities of EV infrastructure, paving the way for scalable and user-centric solutions.
\end{abstract}

%%%%%%%%%%%%%%%%%%%%%%%%%%%%%%%%%%%%%%%%%%%%%%%%%%%%%%%%%%%%%%%%%%%%%%%%%%%%%%%%
\section{INTRODUCTION}

The adoption of electric vehicles (EVs) is rapidly increasing, driven by a growing awareness of environmental issues and the need to reduce carbon emissions to fight climate change. Governments are offering incentives such as subsidies, tax breaks, and significant investments in charging infrastructure, making the transition to EVs more appealing and practical for both individuals and businesses. This rise in EV usage is transforming the transportation sector, leading to a greater demand for a well-developed and accessible charging network to accommodate the increasing number of EV users~\cite{liu2012optimal,ye2022learning}.

Given the increase in the number of EVs, many charging stations have been established across various regions in cities. However, some stations experience heavy demand during the week, while others remain underutilized. It is crucial to implement a smart pricing strategy that can help balance the network in real time. By reducing the price at low-demand stations, EV drivers can be encouraged to charge their vehicles at less crowded locations. This strategy effectively distributes the charging load across the network and helps avoid traffic congestion. This not only improves user convenience by reducing wait times and ensuring access to charging but also benefits service providers by optimizing the utilization of their infrastructure, ultimately leading to a more efficient and balanced charging ecosystem ~\cite{feng2020review,ai2018household,su2022operating}.

There is a rich body of research on EV charging infrastructure and the associated economics. Early studies have shown that charging prices are one of the critical determinants for users when selecting charging stations \cite{hu2016pricing}. Dynamic pricing strategies have been explored extensively in recent years, with studies emphasizing their ability to manage energy load and optimize the performance of charging stations. Authors in \cite{kalakanti2024dynamic} highlight the limitations of traditional pricing mechanisms, such as time-of-use (ToU) rates, which fail to adapt to real-time demand fluctuations. Their research proposes a dynamic pricing model addressing multiple conflicting objectives, including revenue generation, quality of service, and peak-to-average ratios, utilizing advanced algorithms like non-dominated sorting genetic algorithms (NSGA)~II and III to find optimal trade-offs. Integrating machine learning and deep learning approaches, such as long short-term memory (LSTM), has also been increasingly investigated to enhance price optimization. Recent work in \cite {kuang2024unraveling} provides a detailed analysis of the effect of electricity prices on EV charging behavior using a learning model incorporating a two-layer graph and temporal pattern attention. In addition to price optimization, demand prediction plays a crucial role in ensuring that charging infrastructure is adequately prepared for fluctuations in demand. Accurate demand prediction models help service providers anticipate peak usage periods and adjust their strategies accordingly~\cite{yang2023}. Machine learning techniques, such as neural networks and gradient boosting, have been widely used for forecasting EV charging demand, providing valuable insights into usage \cite{sanami2025, majidpour2016forecasting, lu2018application, lu2017load}.

Although dynamic pricing has been studied, there remains a knowledge gap in addressing network balancing that takes into consideration overutilization and underutilization of charging stations, especially in densely packed urban areas where the interaction between multiple stations is crucial. This paper introduces a novel graph-based reinforcement learning-based approach to optimize network balancing in EV charging stations. The primary objective of this research is to enhance load distribution by dynamically adjusting pricing strategies in near real time. To achieve this, we represent the charging station network as a graph. By formulating the problem as a deep Q-learning (DQL) task, the model leverages graph neural networks (GNN) to understand interdependencies between stations, addressing the challenges of optimizing price elasticity to balance demand, reduce over-utilization, and improve overall network efficiency. The DQL-based model learns an optimal pricing strategy that dynamically adapts to changes in charging demand and reduces overload events.

The organization of the paper is as follows. Section~\ref{sec:problem} defines the preliminaries and presents the problem statement. The proposed method is discussed in Section~\ref{sec:method}. The implementation of the method and simulation results are outlined in Section~\ref{sec:exp}. Finally, Section~\ref{sec:conc} offers the conclusion of the paper.

\section{Preliminaries and Problem Statement}
\label{sec:problem}
In a standard RL framework, an agent learns to maximize cumulative reward through interactions with an environment. RL problems are commonly modelled as Markov decision processes (MDPs), which are defined by a tuple \( (S, A, T, R) \). Here, \( S \) represents the set of possible states of the environment, \( A \) denotes the set of actions available to the agent, \( T \) is the state transition function describing how actions affect future states, and \( R \) is the reward function that evaluates the desirability of each state-action pair. At each time step \( t \), the agent observes the current state \( s_t \in S \), selects an action \( a_t \in A \), and receives a reward \( R(s_t, a_t) \) from the environment. The objective of RL is to learn a policy \( \pi(a|s) \) that maximizes the expected cumulative reward, known as the return, defined as:
\begin{equation*}
    G_t = \sum_{k=0}^{\infty} \gamma^k R_{t+k+1}
\end{equation*}
where \( \gamma \in [0,1] \) is a discount factor that controls the importance of future rewards relative to immediate rewards.

In this work, we employ deep Q-learning (DQL), a value-based RL approach that leverages deep neural networks to approximate the Q-value function \( Q(s, a) \). The Q-value function estimates the expected return of taking action \( a \) in state \( s \) and following the policy thereafter. The optimal Q-value function \( Q^*(s, a) \) satisfies the Bellman equation:
\begin{equation*}
    Q^*(s, a) = \mathbb{E} \left[ R(s, a) + \gamma \max_{a'} Q^*(s', a') \mid s, a \right],
\end{equation*}
where the agent learns this function to derive an optimal policy \( \pi^*(s) = \arg \max_a Q^*(s, a) \). To improve stability, particularly in large, complex environments, DQL incorporates experience replay and a target network. Experience replay stores past experiences in a buffer and samples them randomly to break the correlation between consecutive experiences, enhancing training stability. A target network, which is updated less frequently, provides a stable reference for Q-value updates, further aiding convergence. Double Q-learning, a variant of DQL, addresses overestimation bias by using two Q-networks, one for action selection and the other for evaluation.

In the current balancing problem in EV charging networks, the RL agent aims to manage charging demand by adjusting prices at each station and directing users to underutilized stations to achieve balanced load distribution. Each charging station is represented as a node in a graph, with edges denoting adjacency between stations. The objective of this problem is to balance the utilization of resources across the network by adjusting prices at individual stations in a way that encourages or discourages demand, thereby achieving equitable load distribution across the network.

The problem formulation leverages the MDP framework, where the state \( s \in S \) represents the real-time status of the network, including the utilization of each station—measured as the ratio of current load to capacity—and the average utilization within the network, reflecting local load distribution. Additionally, the state incorporates demand projections derived from a pre-trained price elasticity model, allowing the agent to forecast the impact of price adjustments at each station. The action space \( A \) consists of possible price adjustments at each station, which influence demand by encouraging users to select underutilized stations and deterring the use of heavily utilized ones. The transition function \( T(s'|s, a) \) accounts for both the spatial dependencies captured by the network structure and the elasticity of demand to price changes, which is modelled through a GNN. The GNN provides spatially-aware demand predictions, enabling the agent to anticipate the effects of localized price adjustments on broader network dynamics.

\section{Methodology}
\label{sec:method}
The primary objective is to ensure balanced utilization across the network, which involves minimizing load disparities among stations and preventing overloads at any one station. By managing utilization levels dynamically, the agent can reduce waiting times and prevent inefficiencies caused by uneven resource distribution. The proposed solution uses a DQL agent, enhanced with a GNN, to make price adjustments that achieve balanced utilization across the EV charging network. The reward function is designed to incentivize balancedness within the network and penalize configurations where any station is near overload.

\subsection{Reward Function Design}

The reward function consists of two primary components: a balancedness term to minimize utilization variance within the network and an overload penalty to discourage near-capacity operation at any station,
% \begin{equation}
% \label{eq:r_cluster}
%     \begin{aligned}
%         R_{\text{cluster}, k} = &-\sum_{i \in \mathcal{C}_k} \left( \frac{L_i}{C_i} - \frac{\bar{L}_{\mathcal{C}_k}}{\bar{C}_{\mathcal{C}_k}} \right)^2 \\
%         &- \lambda \sum_{i \in \mathcal{C}_k} \frac{1}{1 + e^{-\kappa \left(\frac{L_i}{C_i} - 0.9\right)}},
%     \end{aligned}
% \end{equation}
\begin{equation}
\label{eq:reward}
    R = \left[\sum_{i =1}^{N} \left( \frac{L_i}{C_i} - \frac{\bar{L}}{\bar{C}} \right)^2 + \lambda \sum_{i =1}^{N}   \frac{1}{1 + e^{-\kappa \left(\frac{L_i}{C_i} - 0.9\right)}}\right]^{-1}
\end{equation}
where $N$ is the total number of stations, and \( C_i \) and \( L_i \) are the capacity and load of station \( i \), i.e., the total number of charging piles and the number of occupied ones at station \( i \), respectively. Also, \( \bar{L}\) and \( \bar{C}\) represent the network-wide averages for load and capacity. The balancedness term in the reward promotes the reduction of variance in utilization, ensuring an even load distribution among stations. The sigmoid penalty function (second term of~\eqref{eq:reward}) remains near zero for utilizations up to 80\% for $\kappa=30$, but sharply increases as utilization approaches full capacity, thus discouraging overloads. Moreover, the inverse reward function effectively incentivizes the agent to minimize penalties by leveraging its steep sensitivity at low penalty values. This formulation ensures that even small reductions in penalties result in significant increases in reward, driving the agent toward near-optimal behavior. Additionally, the nonlinear scaling provided by the inverse reward formulation enables the agent to prioritize improvements where they matter most, avoiding disproportionate attention to penalties that are already large while encouraging consistent optimization across all penalty types.

% To promote network-wide balancedness, the cluster-level rewards are aggregated into a global reward function, weighted by each cluster's capacity. This global reward \( R_{\text{global}} \) is defined as:
% \begin{equation}
%     R_{\text{global}} = \frac{1}{\sum_{k} C_{\mathcal{C}_k}} \sum_{k} C_{\mathcal{C}_k} \cdot R_{\text{cluster}, k},
% \end{equation}
% where \( C_{\mathcal{C}_k} = \sum_{i \in \mathcal{C}_k} C_i \). The weighted aggregation ensures that larger clusters, which handle more demand, exert a proportionally greater influence on the global balancedness objective.

\subsection{Deep Q-Learning with GNN for Demand Prediction}

% The DQL agent learns an optimal pricing policy by iteratively updating a Q-network that approximates the expected reward for each state-action pair. The agent's training process includes experience replay, which stores past interactions in a buffer and samples them randomly to reduce the correlation between consecutive updates, improving learning stability. A target network, updated at a slower rate, provides stable Q-value targets during training updates.

% The GNN model is used to provide spatially-aware demand predictions, capturing the impact of price adjustments at each station on neighboring stations. By incorporating elasticity estimates into the GNN, the model adjusts demand projections based on the anticipated response to price changes, enabling the agent to make spatially-informed decisions. This DQL-GNN framework thus provides adaptive, data-driven pricing strategies that dynamically balance EV charging load across a complex network, achieving equitable utilization across clusters and preventing overloads in response to varying demand patterns.

The proposed DQL framework integrates a pre-trained Price-Adjusted Graph Neural Network (PAG) model~\cite{kuang2024unraveling} with a Multi-Layer Perceptron (MLP) to optimize electric vehicle (EV) charging prices across a network of charging stations. Unlike traditional Q-learning approaches with simplistic environment simulations, this method leverages the PAG model to predict price elasticity, capturing the spatial and temporal interdependencies of charging demand across stations. This enables the agent to consider the effects of pricing decisions on network-wide utilization.

The PAG model acts as the environment step, predicting demand adjustments based on the current state of occupancy and pricing. The output of the PAG model, which represents price-adjusted station utilization, is fed into an MLP to form the Q-network. The Q-network consists of two fully connected layers and a final output layer that maps to the Q-values for all possible state-action pairs. This design allows the agent to predict expected rewards for pricing actions, supporting informed decision-making to minimize penalties and maximize network efficiency.

Training involves standard DQL techniques, including experience replay and a target network. Experience replay buffers past transitions for randomized sampling, reducing the correlations between updates and stabilizing the learning process. The target network, updated periodically, provides fixed Q-value targets, further enhancing training stability. The reward function encourages minimizing variance and overload penalties, promoting balanced utilization across stations while discouraging excessive demand that could lead to capacity constraints.

This approach enables adaptive, data-driven pricing strategies that respond dynamically to changing demand patterns across the network. By integrating price elasticity predictions into the reinforcement learning process, the proposed framework ensures efficient resource allocation, equitable station utilization, and the prevention of overloads, creating a robust solution for managing EV charging demand in large charging networks.

\section{Experimental testing of the method}
\label{sec:exp}
\subsection{Dataset}
\begin{figure}
    \centering
    \includegraphics[width=.85\linewidth]{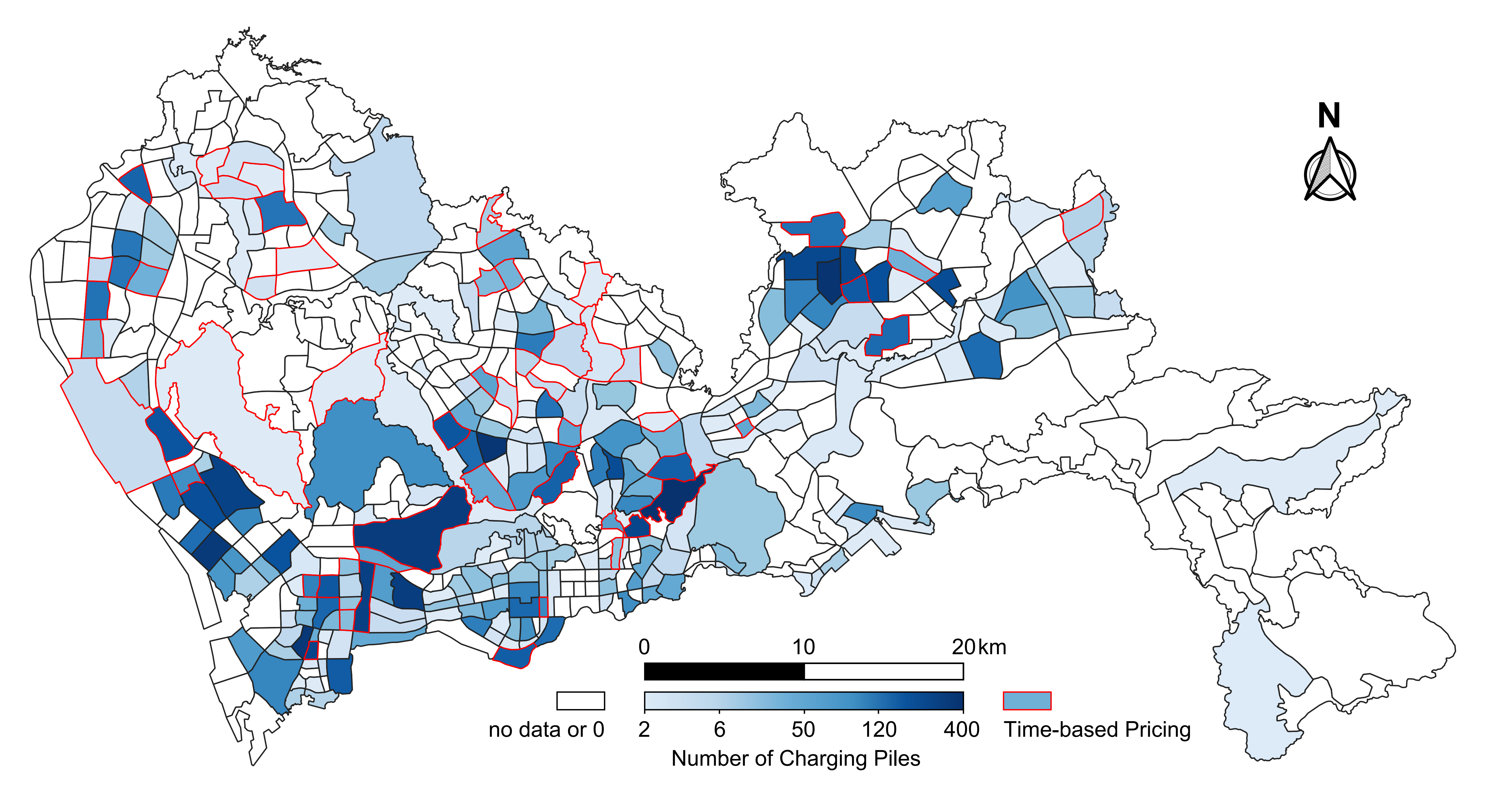}
    \caption{Spatial distribution of the public EV charging piles in ST-EVCDP~\cite{qu2024physics}}
    \label{fig:density}
\end{figure}

The data used in this study is from the open-source ST-EVCDP dataset~\cite{stevcdp} gathered from a publicly available mobile application that provides real-time status of charging pile availability (i.e., whether they are idle or in use). The dataset covers 18,061 public charging piles in Shenzhen, China, collected over one month from June 19 to July 18, 2024, with a data collection interval of 5 minutes, resulting in 8,640 timestamps. The spatial distribution of the EV charging piles in 491 different traffic regions of the city is depicted in Fig.~\ref{fig:density}. To simplify the analysis of the problem, all charging piles in each region are assumed to be consolidated at the centroid of that region. Therefore, each region is referred to as a station from here on. Moreover, by assigning each region with at least one pile to a node, the city's charging stations network can be represented as a graph-structured data set consisting of 247 nodes. In this graph, two nodes are connected if their corresponding regions are geographically adjacent.

\begin{figure}
    \centering
    \includegraphics[width=.85\linewidth]{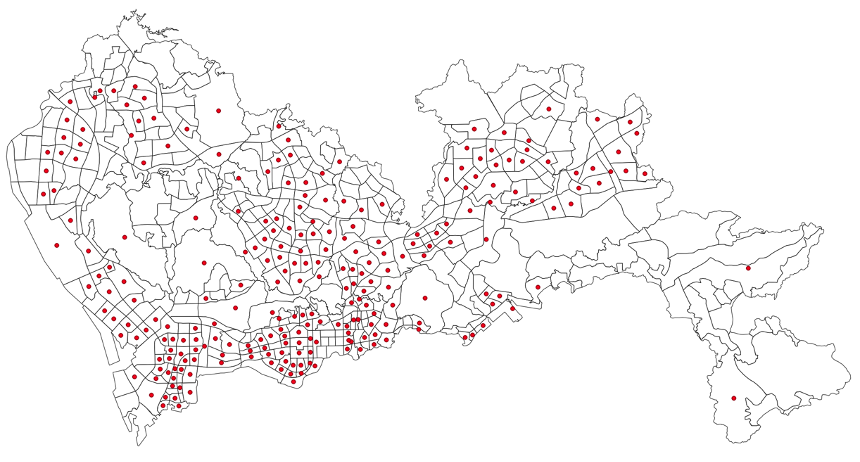}
    \caption{Centroids of the traffic zones with at least one EV charging pile}
    \label{fig:nodes}
\end{figure}

In addition to the geographical specifications of the regions, the ST-EVCDP also includes the occupancy and price records of all stations. The centroid nodes are shown in Fig.~\ref{fig:nodes}.

% \begin{figure}
%     \centering
%     \includegraphics[width=1\linewidth]{density.png}
%     \caption{Density of EV charging piles in different regions of Shenzhen}
%     \label{fig:sensity}
% \end{figure}

\subsection{Data Preprocessing}

To prepare the data for applying the proposed RL method, the raw occupancy and pricing data, initially recorded at 5-minute intervals, are averaged to create 1-hour interval data. While demand exhibits small fluctuations over time, pricing adjustments are typically updated at intervals exceeding two hours. By aggregating the data into 1-hour intervals, we align the temporal granularity with the decision-making needs of the RL framework, reducing noise and improving the efficiency of training without losing the essential long-term patterns in demand and pricing.

In the original geographical configuration shown in Fig.~\ref{fig:nodes}, some regions lack charging piles, leading to a fragmented graph where certain areas are isolated. This disconnected graph structure poses challenges for GNN-based models, which rely on spatial relationships to model demand interactions across neighboring stations. A fully connected graph is crucial to ensure that spatial dependencies are accurately captured and propagated, enabling the RL agent to make informed and globally effective decisions. Without addressing these disconnections, the agent's ability to understand and optimize system-wide behavior would be limited.

\begin{figure}
    \centering
    \includegraphics[width=.85\linewidth]{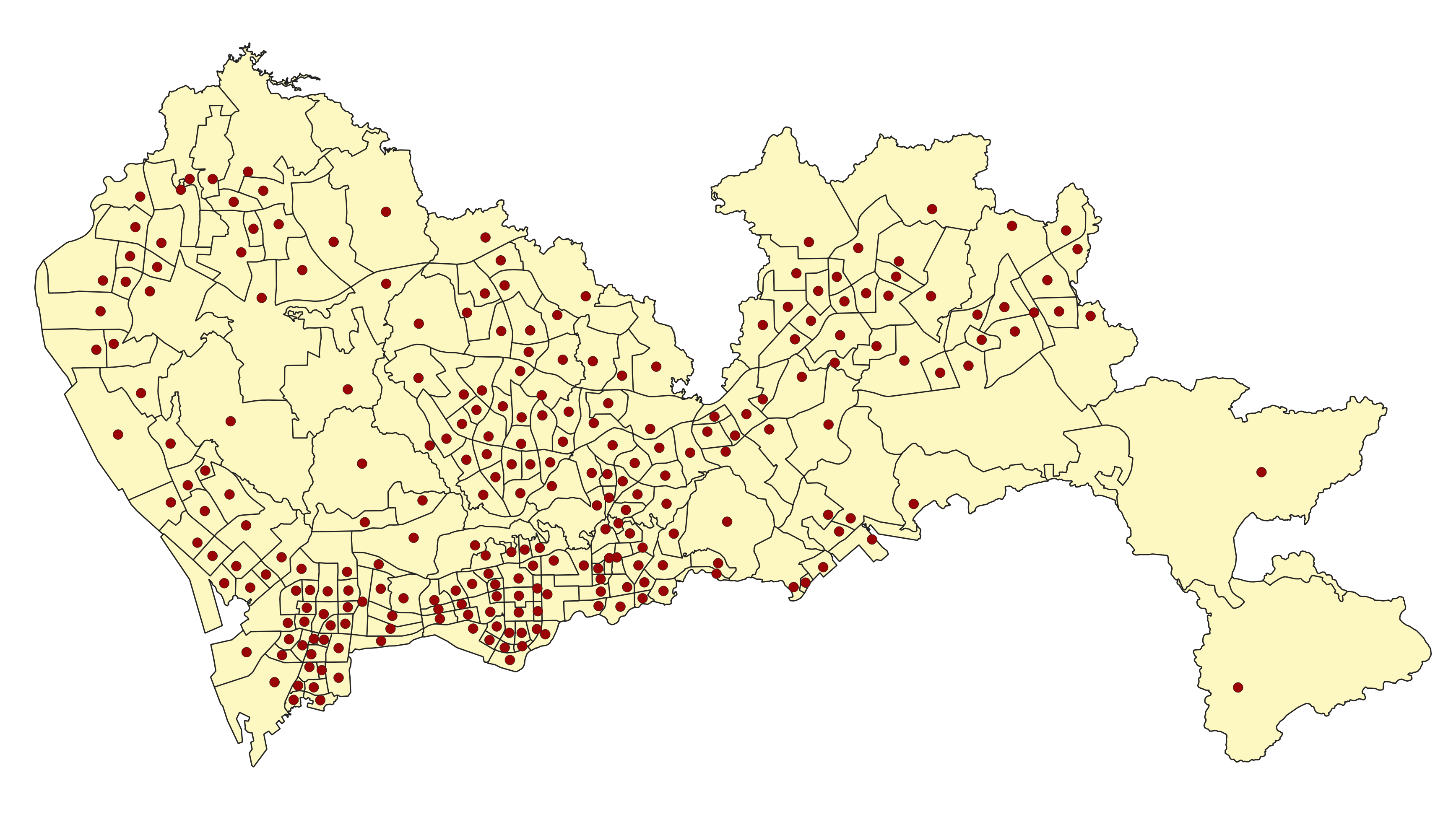}
    \caption{Merged traffic zones map and the centroid nodes of the original stations}
    \label{fig:merged}
\end{figure}

\begin{figure}
    \centering
    \includegraphics[width=.85\linewidth]{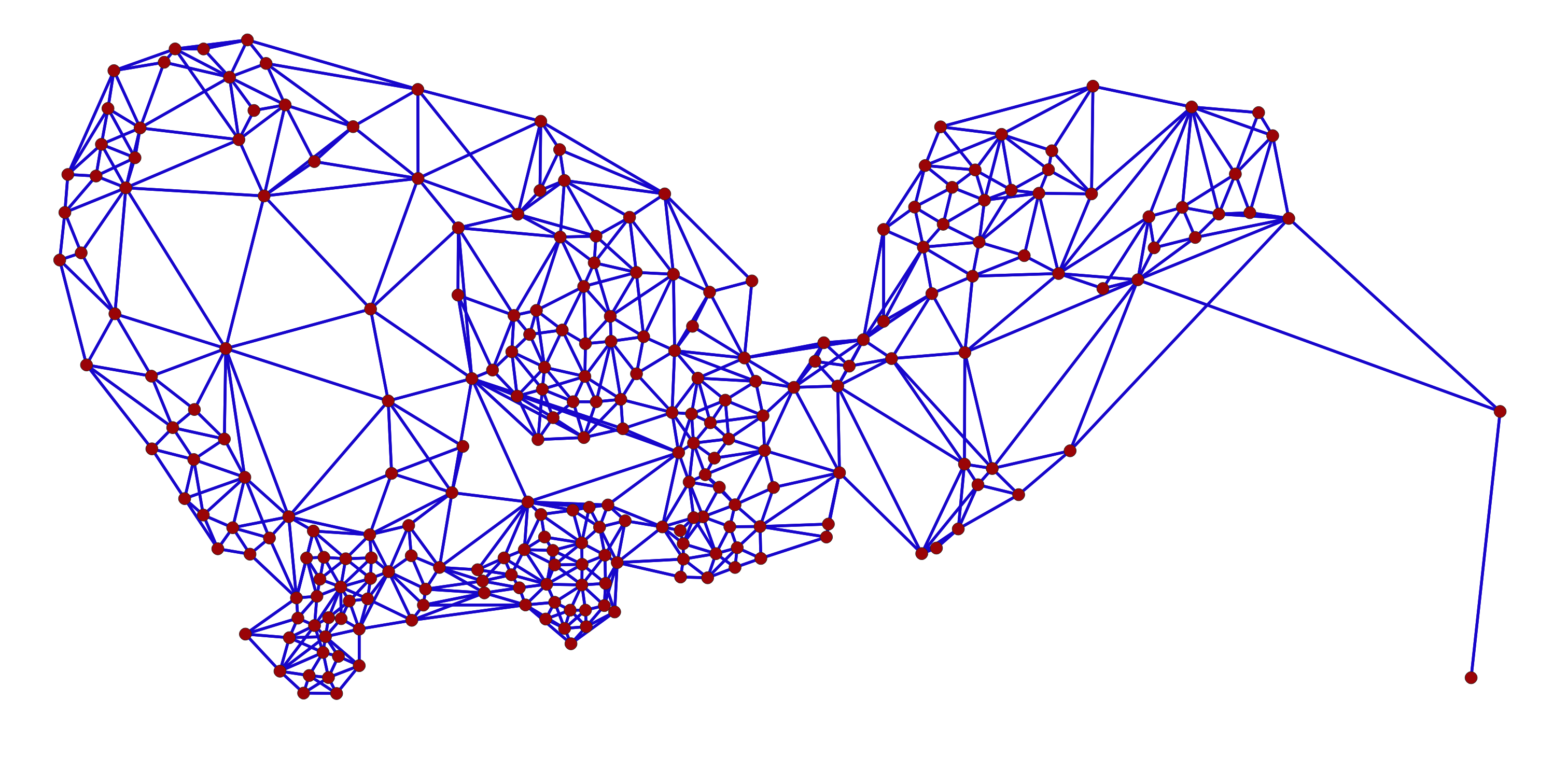}
    \caption{Geographical adjacency graph of the EV charging stations}
    \label{fig:graph}
\end{figure}

To resolve this issue, regions without charging piles are merged with their nearest zone containing at least one charging pile. This merging ensures that the graph representing stations and their adjacencies becomes fully connected. The merging process is performed based on geographical proximity, maintaining realistic spatial relationships while enabling the GNN to capture spatial dependencies across the entire network effectively. This ensures the resulting graph is both meaningful and computationally feasible for modelling. The outcome of this preprocessing step and the resulting graph are illustrated in Figs.~\ref{fig:merged} and~\ref{fig:graph}, respectively. These preprocessing steps ensure the spatial interdependencies among stations are accurately represented, allowing the proposed framework to balance demand across the network effectively.


\subsection{Simulation Results}

The proposed RL approach is evaluated through a series of simulations conducted using the processed EV charging demand dataset. The RL model is trained in a data-driven manner, where each episode processes all training data in mini-batches. This setup allows the agent to refine its policy across multiple episodes iteratively. At the start of training, the agent explores various actions due to the high epsilon value (\(\epsilon = 1.0\)). As training progresses, epsilon decays, encouraging the agent to exploit learned policies while maintaining some level of exploration.

The action space in the proposed RL model consists of price adjustments that each charging station can apply within a predefined range. According to~\cite{kuang2024unraveling}, the average price per kWh across all stations is approximately 0.99 CNY/kWh during daytime and 0.93 CNY/kWh during nighttime. The minimum price observed is 0.54 CNY/kWh, and the maximum price is 1.47 CNY/kWh. These statistics provide the context for the action space, where each action represents a relative price change within the range of \{-0.3, -0.2, -0.1, 0, 0.1, 0.2, 0.3\}, allowing the RL agent to adjust station-specific pricing dynamically.

During training, the Q-network is updated using the Bellman equation, where the target Q-value is computed from the reward and the estimated future Q-value of the next state. In this data-driven framework, the absence of a temporal sequence of state transitions within each batch differs from the traditional interpretation of the Q-value. Instead of reflecting long-term reward expectations over an episode, the Q-value in this setup approximates the immediate cumulative reward associated with the current state-action pair, as derived from the training data. The simulation parameters used in training the RL model are summarized in Table~\ref{tab:params}.

\begin{figure}[!h]
    \centering
    \includegraphics[width=.8\linewidth]{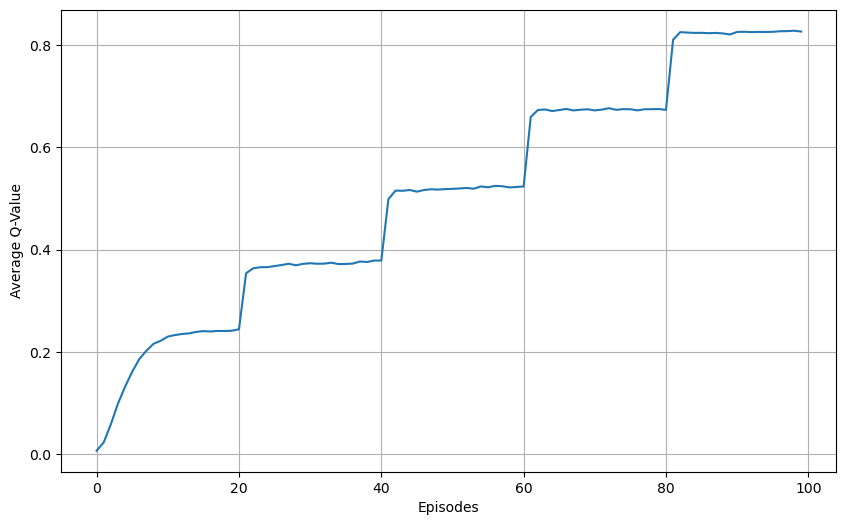}
    \caption{Average Q-Values over episodes}
    \label{fig:q_values}
\end{figure}

\begin{figure}[t]
    \centering
    \includegraphics[width=.88\linewidth]{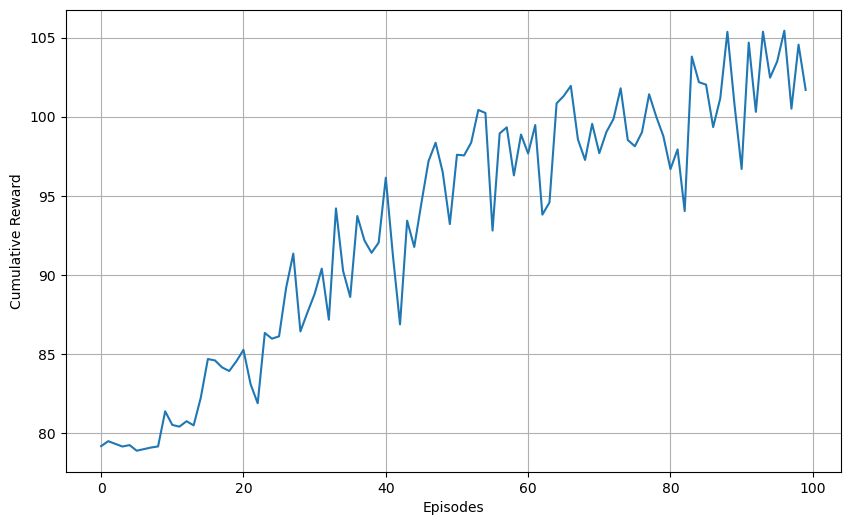}
    \caption{Cumulative rewards over episodes}
    \label{fig:rewards}
\end{figure}

Since the reward function defined in~\eqref{eq:reward} is designed as a reciprocal function of penalties, it is not upper-bounded, as reducing penalties to near-zero results in unbounded growth of the reward, unlike typical bounded reward functions. As a result, the Q-values shown in Fig.~\ref{fig:q_values} tend to rise as the agent improves its policy, indicating advancements in both system balance and load distribution. Moreover, the target network is updated every 20 episodes to stabilize the training process by providing fixed Q-value targets over multiple updates. This results in the step-like increases observed in the Q-value graph (Fig.~\ref{fig:q_values}). The increase in cumulative reward values, as depicted in Fig.~\ref{fig:rewards}, shows that the agent is successfully learning to optimize its policy.

\begin{figure}[!h]
    \centering
    \includegraphics[width=.8\linewidth]{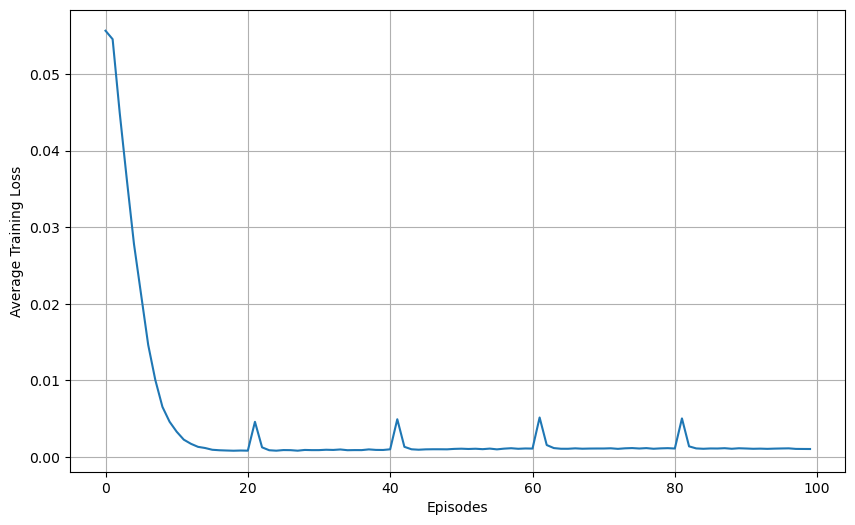}
    \caption{Average training loss over episodes}
    \label{fig:loss_plot}
\end{figure}

The training loss curve, presented in Fig.~\ref{fig:loss_plot}, illustrates the learning dynamics of the Q-network over episodes. The rapid reduction in loss during the initial episodes reflects the agent's effective adaptation to the environment and convergence toward meaningful policy updates. Subsequent oscillations in the loss are attributed to target network updates, which periodically shift the optimization objective.

\begin{figure}[t]
    \centering
    \includegraphics[width=.9\linewidth]{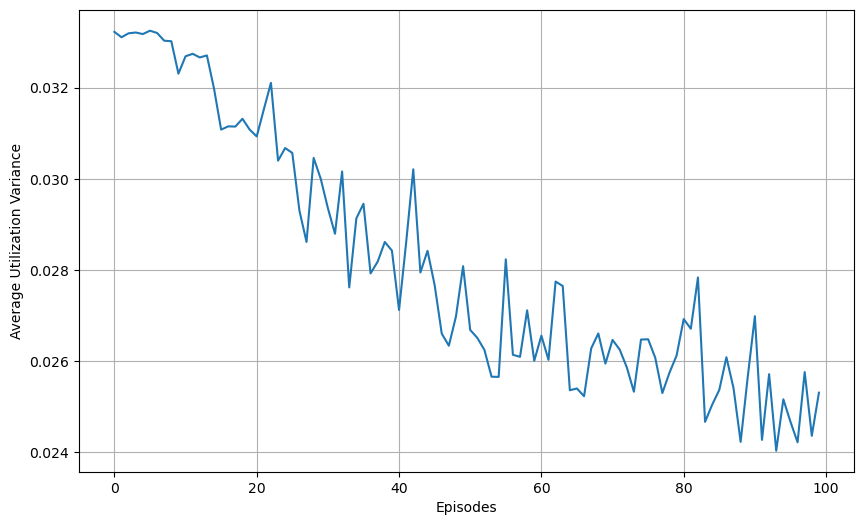}
    \caption{Average variance of utilization over episodes}
    \label{fig:variance}
\end{figure}

Additionally, the average variance of the utilization, shown in Fig.~\ref{fig:variance}, highlights the agent's success in minimizing imbalances in utilization across stations while preventing overloading and underutilization across the charging network.

\begin{figure}[t]
    \centering
    \includegraphics[width=.9\linewidth]{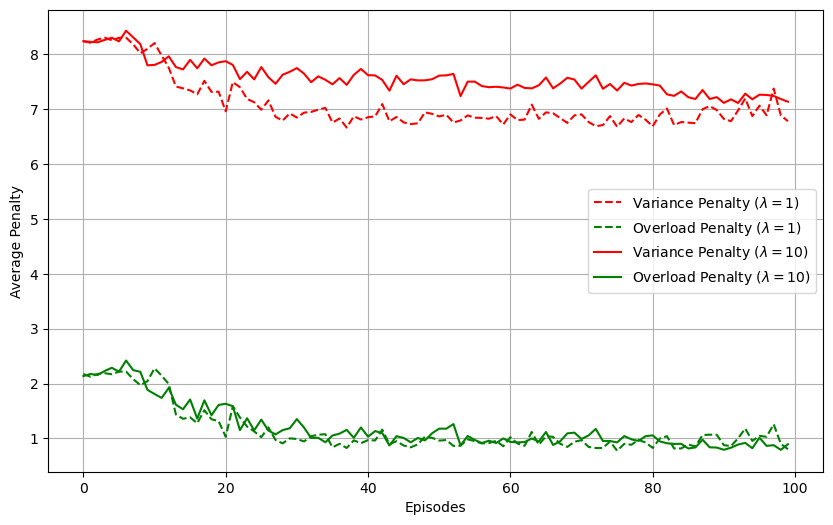}
    \caption{Comparison of penalty components for $\lambda=1$ and $\lambda=10$.}
    \label{fig:penalty_comparison}
\end{figure}

The penalty comparison for two different weightings of the overload penalty ($\lambda=1$ and $\lambda=10$) is depicted in Fig.~\ref{fig:penalty_comparison}. The results highlight the trade-off between variance and overload penalties as $\lambda$ is adjusted. A higher $\lambda$ shifts the balance of the reward function toward addressing overload penalties. Specifically, this prioritization results in higher variance penalties due to imbalances in utilization across stations. These observations demonstrate the flexibility of the proposed framework in adapting to varying operational objectives by tuning the $\lambda$ parameter.

\begin{table}[h]
    \caption{Simulation Parameters}
    \label{tab:params}
    \centering
    \begin{tabular}{|l|c|}
        \hline
        \textbf{Parameter} & \textbf{Value} \\
        \hline
        Discount factor (\(\gamma\)) & 0.99 \\
        Learning rate & \(1 \times 10^{-4}\) \\
        Batch size & 32 \\
        Target network update frequency & 20 episodes \\
        Epsilon decay factor & 0.95 \\
        Minimum epsilon (\(\epsilon_{\text{min}}\)) & 0.1\\
        Overload penalty weight \(\lambda\) & 1\\
        Number of episodes & 100\\
        \hline
    \end{tabular}
\end{table}

\section{Conclusion}
\label{sec:conc}

This paper proposes a reinforcement learning-based framework for dynamic pricing and load balancing in EV charging networks. By leveraging a pre-trained price-adjusted graph neural network (PAG) for demand prediction and a reward function designed to minimize both utilization variance and overload penalties, the framework demonstrates its ability to optimize resource allocation and achieve equitable station utilization. The simulation results validate the efficacy of the proposed approach, with significant improvements observed in balancing network demand and preventing capacity overloads across stations. Additionally, the flexibility of the reward function is highlighted through its adaptability to varying operational priorities by adjusting the penalty weighting factor \(\lambda\). Future work will focus on extending this framework to incorporate real-time constraints, such as power resource management and user preferences, further enhancing the applicability of the proposed solution to real-world scenarios.

\bibliographystyle{IEEEtran} % This specifies the style in which the references are formatted. Popular styles include `plain`, `unsrt`, `alpha`, and `abbrv`.

\bibliography{reference}

\end{document}